\documentclass[twocolumn,aps,prc,superscriptaddress]{revtex4}
\usepackage{amssymb}
\usepackage{amsmath}
\usepackage{graphicx}
\usepackage[normalem]{ulem}
\usepackage{multirow}
\usepackage{appendix}
\usepackage{CJK}
\usepackage[usenames]{color}
\usepackage{bm}
\begin{document}
\title{Simulating the Chiral Magnetic Wave in a Box System}
\author{Wen-Hao Zhou}
\affiliation{Shanghai Institute of Applied Physics, Chinese Academy of Sciences, Shanghai 201800, China}
\affiliation{University of Chinese Academy of Sciences, Beijing 100049, China}
\author{Jun Xu\footnote{correspond author: xujun@sinap.ac.cn}}
\affiliation{Shanghai Advanced Research Institute, Chinese Academy of Sciences, Shanghai 201210, China}
\affiliation{Shanghai Institute of Applied Physics, Chinese Academy of Sciences, Shanghai 201800, China}

\begin{abstract}
    The chiral magnetic wave from the interplay between the chiral magnetic effect and the chiral separation effect is simulated in a box system with the periodic boundary condition based on the chiral kinetic equations of motion. Simulation results are compared with available limits from theoretical derivations, and effects of the temperature, the magnetic field, and the specific shear viscosity on the key properties of the chiral magnetic wave are discussed. Our study serves as a baseline for further simulations of chiral anomalies in relativistic heavy-ion collisions.
\end{abstract}

\maketitle

    \section{Introduction}

    Phenomena induced by chiral anomalies are of great interest in not only material science~\cite{nature,science,PRX} but also heavy-ion physics~\cite{Kha16,Hua16}. The extremely high energy density, at which the chiral symmetry for quarks is restored, and the strong magnetic field can be produced in non-central relativistic heavy-ion collisions, providing the possibility of investigating various chiral anomaly phenomena. For example, the chiral magnetic effect (CME)~\cite{K2008,K_CME,Fuku1} induced by the non-zero axial charge density can lead to the net electric charge current along the direction of the magnetic field, and the electric charge separation at RHIC and LHC energies can be measured indirectly by comparing the correlation between particles of the same and opposite electric charge~\cite{STAR1,STAR2,ALICE1}. The chiral separation effect (CSE)~\cite{CSE1,CSE2} is a dual effect of the CME induced by non-zero electric charge density, leading to the net axial charge current along the direction of the magnetic field, and may have relevant consequences in the evolution and observables of neutron stars. Since fluctuations of the electric (axial) charge density lead to the net axial (electric) charge current and thus locally the non-zero axial (electric) charge density, the interplay between the CME and the CSE results in a collective excitation called the chiral magnetic wave (CMW)~\cite{K_CMW}, which may generate an electric quadrupole moment in a quark matter and can be responsible for the elliptic flow splitting between particles of opposite electric charges~\cite{elec,STAR15}.

    Recently, the CMS Collaboration found that the electric charge separation signals in p+Pb collisions are similar to that in Pb+Pb collisions~\cite{CMS1,CMS2}, challenging the interpretation that these signals are from the CME. On the other hand, it was found that resonance decays can largely account for the three-particle correlator as the charge separation observable in the search for the CME~\cite{back}. Also, the elliptic flow splitting between particles of opposite electric charges may be originated from their different mean-field potentials~\cite{Xu12,Xu14,Liu16} rather than the CMW effect. In order to remove possible background contributions to the signals from chiral anomalies, new experimental observables are proposed~\cite{Zha17,Mag18,Xu18}. Collisions of isobaric nuclei proposed in the STAR run schedule~\cite{rhic_iso} could be useful to disentangle the signals of chiral anomalies from the background. On the theoretical side, quantitative results from anomalous simulations are very much needed. Based on the hydrodynamic framework, various approaches have been developed by incorporating chiral anomalies~\cite{Hydro1,Hydro2,Liao}. Using the chiral kinetic equations of motion developed in Refs.~\cite{CKM1,CKM3,CKM4}, observables from chiral anomalies in relativistic heavy-ion collisions have been studied~\cite{sun1,sun2} based on the anomalous transport approach, from a blast-wave-like initial condition and with a parameterized decaying magnetic field.

    Toward an accurate description of chiral anomalies with the transport approach, in the present study we simulate the CMW from the interplay between the CME and the CSE in a cubic box system with the periodic boundary condition under a uniform and constant external magnetic field. The simulation in the box system is under well controlled condition and can be compared with limiting results from theoretical derivations, so it has the advantage of calibrating various components of the transport approach to be carried out in a more realistic but complicated heavy-ion system. We note in previous studies the box simulation method has been used to understand the thermalization of gluons~\cite{stochastic}, the specific shear viscosity of the relativistic hadron gas~\cite{Dem09}, equilibrium properties of nucleons and pion-like particles~\cite{Zha18}, and the collision rate of nucleon-nucleon scatterings from various transport codes~\cite{box2}. With the help of the box calculation method, we found that the propagating velocity of the CMW and the development of the electric quadrupole moment are enhanced with the increasing strength of the magnetic field but suppressed with the increasing temperature, while the damping of the CMW is weaker with a smaller specific shear viscosity and/or at a higher temperature.

    This paper is organized as follows. Section \ref{sec:THEORY} briefly reviews the chiral kinetic equations of motion as well as the CME, CSE, and CMW formulisms, and provides the detailed simulation method in a box condition. Section \ref{sec:RESULTS} discusses the detailed CMW results from the interplay between the CME and the CSE. A summary is given in Sec. \ref{sec:SUMMARY}.

    \section{Theoretical framework}
    \label{sec:THEORY}
    \subsection{Chiral kinetic equations of motion}

    In the following we derive intuitively the chiral kinetic equations of motion (EOMs) for massless spin$-1/2$ particles under an external magnetic field in a semiclassical way. In the massless limit, the Hamiltonian for particles with helicity $c$ can be written as %\cite{Massive}
    \begin{equation}
    H = c\vec{\sigma} \cdot \vec{k} \label{Ham}
    \end{equation}
    with
    \begin{equation}
    \vec{k} = \vec{p} - qe\vec{A}.
    \end{equation}
    In the above, $\vec{\sigma}$ is the Pauli matrix, $\vec{k}$ and $\vec{p}$ are respectively the kinematic and canonical momentum, $q$ is the charge number of the particle, and $\vec{A}$ is the vector potential with $\vec{B} = \nabla \times \vec{A}$ being the magnetic field.

    Treating $\vec{r}$, $\vec{k}$, and $\vec{\sigma}$ as classical variables, the canonical EOMs obtained from Eq. (\ref{Ham}) can be written as
    %-------------------------------------% r,k,sigma evolution
    \begin{gather}
    \dot{\vec{r}} = c\vec{\sigma}, \label{spin1}\\
    \dot{\vec{k}} = c\vec{\sigma} \times qe\vec{B},\label{spin2}\\
    \dot{\vec{\sigma}} = c\frac{2}{\hbar}\vec{k} \times \vec{\sigma}.\label{spin3}
    \end{gather}
    Using the adiabatic approximation that the spin $\vec{\sigma}$ of the particle evolves much faster than $\vec{r}$ and $\vec{k}$, $\vec{\sigma}$ can be expanded in the first order as \cite{Huang,huangy}
    \begin{equation}
    \vec{\sigma} \approx c\hat{k} - \frac{\hbar}{2k}\left( \hat{k} \times \dot{\hat{k}} \right), \label{appr}
    \end{equation}
    where $\hat k=\vec{k}/k$ is a unit vector. The above expansion is important in obtaining the chiral kinetic EOMs, but also leads to the unavoidable problem that the whole framework is not valid for particles with too small momenta $k$ (see Ref.~\cite{CKM1} for more detailed discussions).

    Substituting Eq.~(\ref{appr}) into Eqs.~(\ref{spin1}-\ref{spin3}), we obtained the chiral kinetic EOMs as in Refs.~\cite{CKM1,CKM3,CKM4}
    \begin{align}
    \sqrt{G}\dot{\vec{r}} &= \hat{k} + \hbar \left( c\vec{b} \cdot \hat{k} \right) qe\vec{B}, \label{chi1} \\
    \sqrt{G}\dot{\vec{k}} &= \hat{k} \times qe\vec{B}, \label{chi2}
    \end{align}
    with
    \begin{equation}
    \sqrt{G}              = 1+\hbar \left( qe\vec{B} \cdot c\vec{b} \right), \label{chi3}
    \end{equation}
    where $\vec{b}\!=\!\vec{k}/\!\left(2k^3\right)$ denotes the so-called Berry curvature \cite{Berry1984}. Since the $\sqrt{G}$ factor in the above EOMs leads to the modification of the phase-space volume or the density of states~\cite{gc}, the phase-space integral $\mathrm{d}^3r\mathrm{d}^3k/\left(2\pi\hbar\right)^3$ needs to be modified to $\sqrt{G}\mathrm{d}^3r\mathrm{d}^3k/\left(2\pi\hbar\right)^3$. In the transport simulation, the average value of an observable is correspondingly calculated according to $\langle A \rangle = \sum_i A_i \sqrt{G_i} / \sum_i  \sqrt{G_i}$ by taking $\sqrt{G}$ as a weight factor.

    \subsection{Formulisms of chiral magnetic wave}

    The charge density and current of particles with the charge number $q$ and the helicity $c$ can be respectively expressed as
    \begin{align}
    \rho_{qc} &= qN_c\int \frac{\mathrm{d}^3k}{\left( 2\pi\hbar \right)^3}\sqrt{G}f\left(\frac{k-\mu_{qc}}{T}\right),\label{rho_4} \\
    \vec{J}_{qc} &= N_c\int \frac{\mathrm{d}^3k}{\left( 2\pi\hbar \right)^3}\sqrt{G}\dot{\vec r}f\left(\frac{k-\mu_{qc}}{T}\right),\label{J_4}
    \end{align}
    where $N_c=3$ is the color degeneracy, $f(x)\!=\!1/\left(e^{x}+1\right)$ represents the Fermi-Dirac distribution, $T$ is the temperature of the system, $\mu_{qc}\!=\!q\mu+c\mu_5$ is the chemical potential of the particle with $\mu\left(\mu_5\right)$ being the electric (axial) charge chemical potential. The electric charge density and current for particles with the right-handed (R) and the left-handed (L) chirality are defined as
    \begin{align}
    \rho_{R} =\rho_{q(+)c(+)} + \rho_{q(-)c(-)}, \label{rho_r}\\
    \rho_{L} =\rho_{q(+)c(-)} + \rho_{q(-)c(+)}, \\
    \vec{J}_{R} = \vec{J}_{q(+)c(+)} - \vec{J}_{q(-)c(-)}, \\
    \vec{J}_{L} = \vec{J}_{q(+)c(-)} - \vec{J}_{q(-)c(+)},
    \end{align}
    and the total electric and axial charge density and current are respectively defined as
    \begin{align}
    \rho = \rho_{R}+\rho_{L}, \quad&\rho_5 = \rho_{R}-\rho_{L}, \\
    \vec{J} = \vec{J}_R+\vec{J}_L, \quad&\vec{J}_5 = \vec{J}_R-\vec{J}_L. \label{J_sim}
    \end{align}

    From an isotropic Fermi-Dirac momentum distribution, Eq. (\ref{J_4}) can be analytically carried out, and the electric and axial charge current become
    \begin{align}
    \vec{J}   &=\frac{N_c}{2\pi^2\hbar^2}\mu_5e\vec{B}, \label{J2}\\
    \vec{J}_5 &=\frac{N_c}{2\pi^2\hbar^2}\mu e\vec{B}.  \label{J52}
    \end{align}
    Equations (\ref{J2}) and (\ref{J52}) represent respectively the CME and the CSE. In the limit of $\mu/T \ll 1$ and $\mu_5/T \ll 1$, the electric and axial charge density are proportional to their corresponding chemical potential
    \begin{align}
    \rho   &\approx \frac{N_cT^2}{3\hbar^3}\mu,   \label{rho}\\
    \rho_5 &\approx \frac{N_cT^2}{3\hbar^3}\mu_5. \label{rho5}
    \end{align}
    Combining Eqs. (\ref{J2}-\ref{rho5}) and using the definitions of Eqs. (\ref{rho_r}-\ref{J_sim}), the following decoupled relation can be obtained
    \begin{equation}
    \vec{J}_{R\!/\!L} = \pm\frac{3\hbar e\vec{B}}{2\pi^2T^2}\rho_{R\!/\!L}, \label{adv}
    \end{equation}
    where the upper (lower) sign is for particles with the right-handed (left-handed) chirality. $\vec{J}_{R\!/\!L}$ has not only the advective contribution as Eq. (\ref{adv}) but also the diffusive term approximated by Fick's law as $-D_L\nabla\rho_{R\!/\!L}$~\cite{K_CMW}, where $D_L$ is the diffusion constant along the direction of the magnetic field, i.e., the $+y$ direction in this study. Considering both the advective and the diffusive term, the continuity equation $\partial_t \rho_{R\!/\!L} + \nabla\cdot\vec{J}_{R\!/\!L}=0$ becomes the following equation describing the CMW~\cite{K_CMW,JHEP}
    \begin{equation}
    \left(\partial_t \pm v_p \partial_y- D_L\partial_y^2\right) \rho_{R\!/\!L} = 0, \label{pequ}
    \end{equation}
    where
    \begin{equation}
    v_p = \frac{3\hbar eB}{2\pi^2T^2} \label{v_p_t}
    \end{equation}
    is the phase velocity. By applying the Fourier series method, the general solution of Eq. (\ref{pequ}) gives the density at position $y$ and time $t$ as
    \begin{align}
    \notag \rho_{R\!/\!L}\!\left(y,t\right)&=\sum_{n=0}^{\infty}\bigg \{ A_ne^{-D_L{\beta_{2n}^2}t}\sin{\left[\beta_{2n}\left(y \mp v_pt\right)\right]} \\
    &+B_ne^{-D_L\beta_{2n+1}^2t}\cos{\left[\beta_{2n+1}\left(y \mp v_pt\right)\right]}\bigg \}, \label{resr}
    \end{align}
    where the coefficients $A_n$ and $B_n$ are related to the initial density distribution $\rho_{R\!/\!L}\!\left(y,0\right)$ through
    \begin{align}
    \notag & A_n = \frac{1}{l}\int_{-l}^{l}\rho_{R\!/\!L}\!\left(y,0\right)\sin{\left(\beta_{2n}y\right)}\mathrm{d}y, \\
    \notag & B_n = \frac{1}{l}\int_{-l}^{l}\rho_{R\!/\!L}\!\left(y,0\right)\cos{\left(\beta_{2n+1}y\right)}\mathrm{d}y.
    \end{align}
    In the above equations, the periodic boundary condition $\rho_{R\!/\!L}|_{y=-l}\!=\!\rho_{R\!/\!L}|_{y=+l}$ is satisfied, and $\beta_m = \frac{m\pi}{2l}$ is the wave number of the CMW.

    \subsection{Description of box simulation}

    The periodic boundary condition, that a particle which goes out of one side of the box enters from the opposite side with the same momentum, is applied to the box simulation. The side length of the cubic box is $2l=10\,\mathrm{fm}$. Massless particles with the charge number $q=1$ and $-1$ are initialized according to the isotropic Fermi-Dirac distribution in momentum space and certain designed distribution in coordinate space, and they evolves according to Eqs. (\ref{chi1}-\ref{chi3}) under a constant and uniform magnetic field in $+y$ direction. As mentioned before, the EOMs are not valid for particles with too small momenta, which may lead to zero or even negative values of $\sqrt{G}$, depending on the strength of the magnetic field. These particles are treated as free ones, and they are not counted in the calculation of the current. Some artificial truncations are applied in the momentum space, and this leads to the underestimate of the CME and the CSE to be discussed later.

    The stochastic method \cite{stochastic} is employed to treat the elastic scattering part of the box simulation. The collision probability for a pair of particles with the energy $E_1$ and $E_2$ in a volume $(\Delta x)^3$ and a time interval $\Delta t$ is
    \begin{equation}
    P_{22} = v_{rel}\sigma_{22}\frac{\Delta t}{(\Delta x)^3},
    \end{equation}
    with $v_{rel}=s/\!\left(2E_1E_2\right)$ where $s$ is the square of the invariant mass of the particle pair. $\sigma_{22}$ is a constant and isotropic cross section, determined by the specific shear viscosity as detailed in \ref{sec:APPENDIX}. The box is divided into small cells with the volume of each cell set as $(\Delta x)^3 = 1$ fm$^3$, and only particle pairs in the same cell can collide with each other. The time step $\Delta t=0.01\,\mathrm{fm/c}$ is used in the simulation. The Fermi-Dirac distribution $f$ in the momentum space is maintained through out the box simulation, since $f$ is used exactly as the Pauli blocking probability for the scatterings between particles.

    \section{RESULTS AND DISCUSSIONS}
    \label{sec:RESULTS}

    In this section, we studied in detail the CMW from the interplay between the CME and the CSE. The limitting results of the electric and axial charge current from the CME and the CSE as well as the phase velocity of the CMW are available for comparison, while other detailed properties of the CMW such as the damping and the electric quadrupole moment can only be reliably obtained from simulations. Dependencies on the temperature, the strength of the magnetic field, and the specific shear viscosity will also be discussed. The simulation results are generally averaged over thousands of events.

    \subsection{Chiral magnetic effect and chiral separation effect}

    \begin{figure}[htb]
\includegraphics[angle=0,scale=0.5]{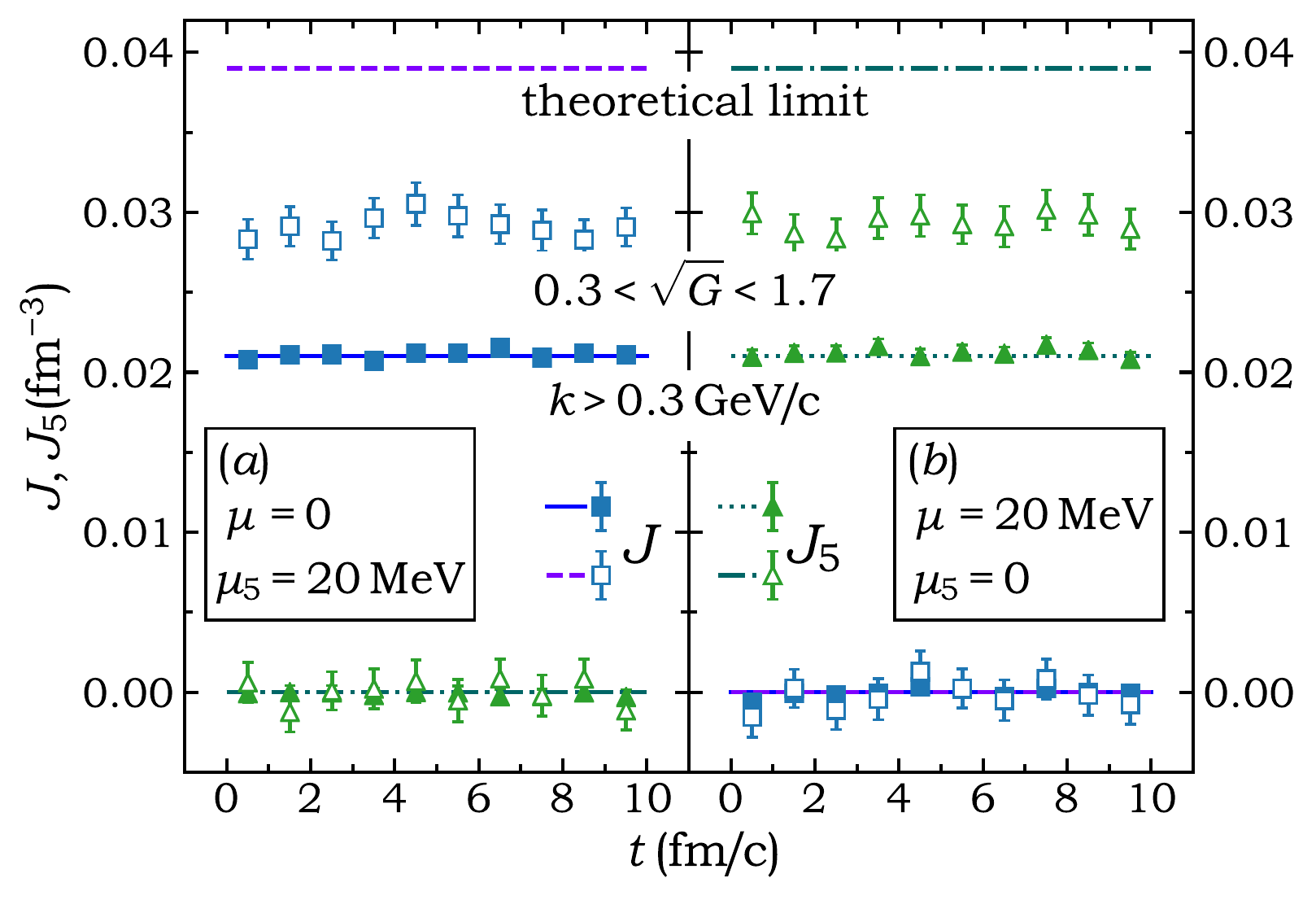}
            \caption{(Color online) Time evolutions of the electric and axial charge current from respectively initial finite axial and electric charge chemical potential, with panel (a) representing the CME and panel (b) representing the CSE. The lines are from theoretical derivations and the symbols are from simulations. Results from the theoretical limit and those from the simulations with truncations of $k>0.3\,\mathrm{GeV/c}$ (solid) and $0.3<\sqrt{G}<1.7$ (open) are compared. See text for more details.}
            \label{J_J5_2}
    \end{figure}

    Before investigating the CMW, we first study separately the CME and the CSE in the box system, by initiating particles with different charge numbers and helicities uniformly distributed in the box with ($\mu=0,\,\mu_5=20\,\mathrm{MeV}$) or ($\mu=20\,\mathrm{MeV},\,\mu_5=0$) at $T=0.3$ GeV, and their momenta sampled according to the isotropic Fermi-Dirac distribution. The strength of the external magnetic field is set as $eB_y=0.5\,\mathrm{GeV/fm}$, and the specific shear viscosity is chosen as $\eta/s=0.08\hbar$. Figure \ref{J_J5_2} displays the electric charge current $\vec{J}$ and the axial charge current $\vec{J}_5$ from theoretical derivations and simulations at different times. The lines represent results from theoretical derivations according to Eqs. (\ref{J_4}) and (\ref{J_sim}), while the symbols represent simulation results from event averaged flux of particles along the magnetic field. It is seen that all currents are constants in time, showing the stability of the simulation. The CSE shown in the right panel is seen to be exactly a dual effect of the CME shown in the left panel. However, with unavoidable truncations in the momentum space to rule out particles with too small momenta, the currents from simulations are always lower than the theoretical limits from Eqs. (\ref{J2}) and (\ref{J52}). Using the truncation of $k_{min}=0.3\,\mathrm{GeV/c}$, the currents from simulations are consistent with available theoretical calculations from Eq. (\ref{J_4}) by integrating from $k_{min}$ to infinity, while the results are only about half of that from the theoretical limit. Using a truncation of $0.3<\sqrt{G}<1.7$ helps to improve the situation, while the corresponding results from theoretical derivations are not available. It is seen that with unavoidable truncations, the strength of the CME and the CSE are underestimated by about $25 \sim 50 \%$ in the simulation. Since the situation from the truncation $0.3<\sqrt{G}<1.7$ is better, it is applied to the investigation of the CMW in the following.

    \subsection{Chiral magnetic wave}

    In order to investigate the CMW, the density distributions are initialized as
    \begin{equation}\label{init}
    \rho_{R\!/\!L}\!\left(y, 0\right) = \pm\frac{1}{2}A_cn\sin{\left(\beta y\right)},
    \end{equation}
    where the upper (lower) sign is for right-handed (left-handed) particles, $n$ is the total average number density in the box system, $A_c\!=\!\left(\rho_R-\rho_L\right)\!/\!n=0.1$ represents the maximum chiral asymmetry in the initial state, and $\beta=\pi/l$ is the chosen wave number. The momenta of particles are sampled isotropically according to the local Fermi-Dirac distribution. The oscillation mode in the later dynamics is determined from the initial condition due to the orthogonality property of trigonometric functions, and the time evolution of the density distributions described by Eq. (\ref{resr}) becomes
    \begin{equation}
    \rho_{R\!/\!L} (y,t) = \pm\frac{1}{2}A_cne^{-D_L{\beta}^2t}\sin{\left[\beta\left(y\mp v_p t \right) \right]}. \label{rhoc}
    \end{equation}
    The electric  and axial charge density can then be expressed as
    \begin{align}
    \rho &= \rho_R+\rho_L = -A_cne^{-D_L{\beta}^2t}\sin{\left(\beta v_p t\right)}\cos{\left(\beta y\right)}, \label{r_charge}\\
    \rho_5 &= \rho_R-\rho_L = +A_cne^{-D_L{\beta}^2t}\cos{\left(\beta v_p t\right)}\sin{\left(\beta y\right)}.
    \end{align}
    \begin{figure}[htb]
\includegraphics[angle=0,scale=0.5]{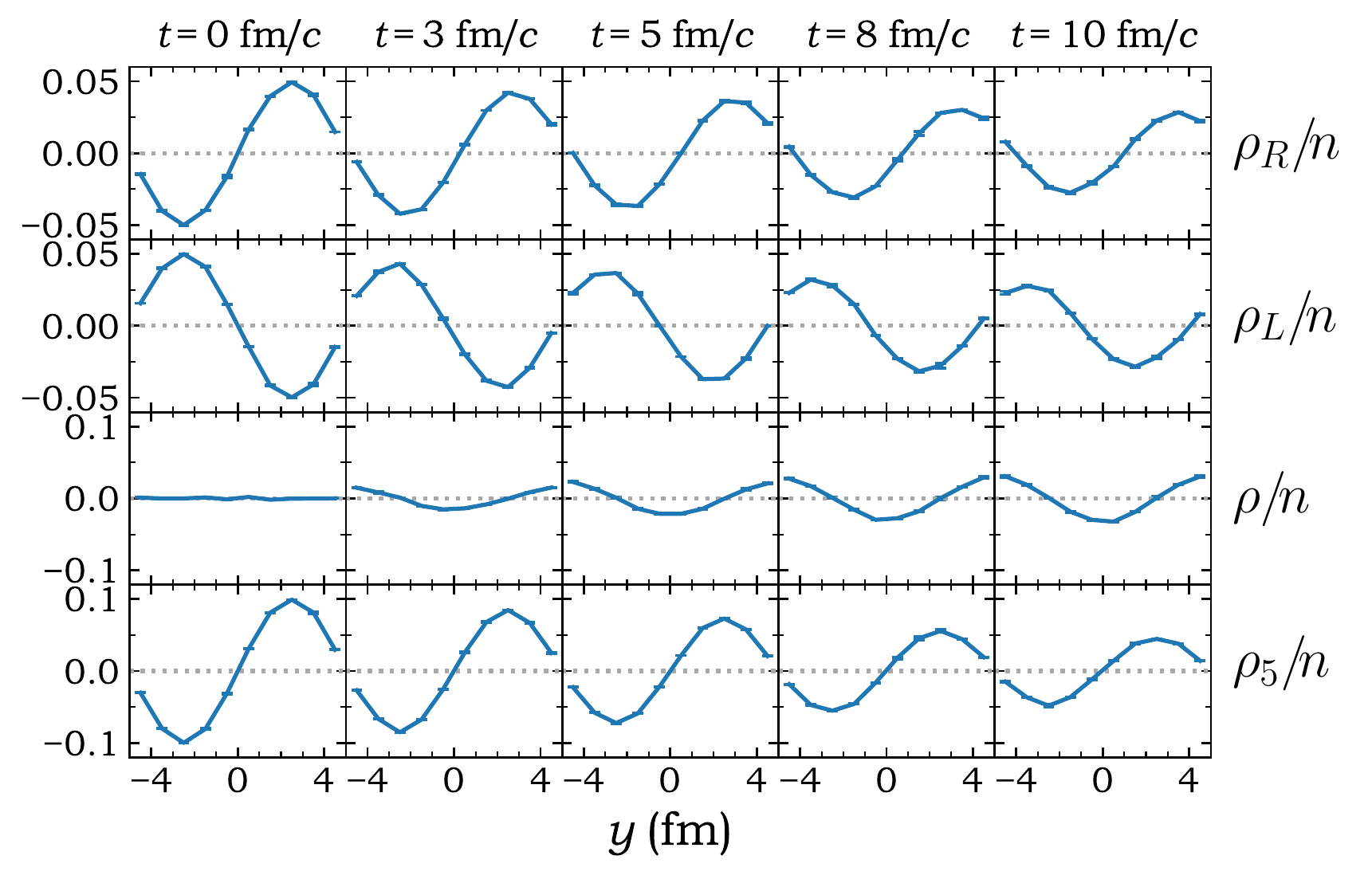}
            \caption{(Color online) Time evolutions of the reduced density of right-handed particles $\rho_R/n$ (first row), the reduced density of left-handed particles $\rho_L/n$ (second row), the reduced electric charge density $\rho/n$ (third row), and the reduced axial charge density $\rho_5/n$ (fourth row) along the direction of the magnetic field from the initial condition of Eq. (\ref{init}). }
            \label{evolve}
    \end{figure}

    With an initial non-uniform axial charge density $\rho_5$ but zero electric charge density $\rho$ described by Eq. (\ref{init}), the time evolutions of various densities are displayed in Fig.~\ref{evolve}. The strength of the magnetic field, the temperature, and the specific shear viscosity are respectively $0.5\,\mathrm{GeV/fm}$, $0.3\,\mathrm{GeV}$, and 0.08$\hbar$, same as in Fig. \ref{J_J5_2}. It is seen that $\rho_R$ ($\rho_L$) is propagating along the $+y$ ($-y$) direction, parallel (antiparallel) to the magnetic field. The dampings of $\rho_R$, $\rho_L$, and $\rho_5$ are observed, while the zero initial electric charge density $\rho$ grows to an electric charge distribution with a quadrupole moment. Since the truncation in the momentum space underestimates the CME and the CSE as seen in Fig. \ref{J_J5_2}, it may also affect the phase velocity of the CMW to be discussed later. The behavior is expected to be similar with an initial non-uniform electric charge density $\rho$ but zero axial charge density $\rho_5$.

    \begin{figure}[htb]
\includegraphics[angle=0,scale=0.5]{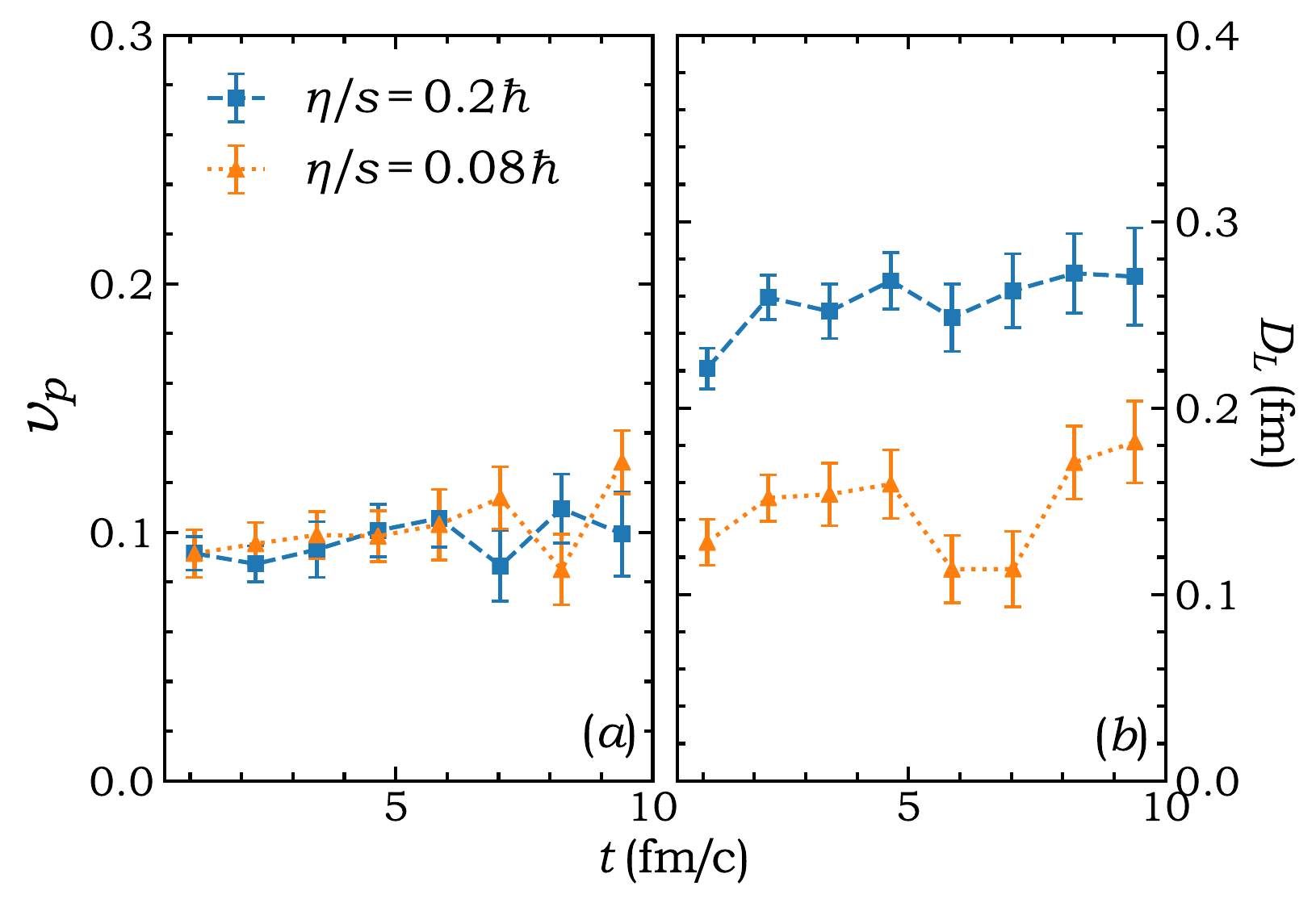}
            \caption{(Color online) The phase velocity $v_p$ (a) and the diffusion constant $D_L$ (b) at different times with different specific shear viscosities $\eta/s$ but the same magnetic field $eB_y=0.5\,\mathrm{GeV/fm}$ and the temperature $T=0.3\,\mathrm{GeV}$.}
            \label{vel_dif}
    \end{figure}

    The phase velocity $v_p$ and the diffusion constant $D_L$ can be obtained by fitting the time evolution of $\rho_R$ and $\rho_L$ shown in the first and the second row of Fig. \ref{evolve} with Eq. (\ref{rhoc}). Their values at different times are shown in Fig. \ref{vel_dif}, and they can be considered approximately as constants in time within the statistical error. It is seen that the phase velocity is insensitive to the specific shear viscosity, while the diffusion constant increases with increasing specific shear viscosity. The later is understandable, since a larger specific shear viscosity corresponds to a smaller cross section and a larger mean free path. Here only one Fourier component of the density fluctuation is considered, while in reality there could be Fourier components with different wave numbers $\beta$. It is expected that higher-order components with larger wave numbers $\beta$ decay faster due to the larger charge density gradient.

    \begin{figure}[htb]
\includegraphics[angle=0,scale=0.5]{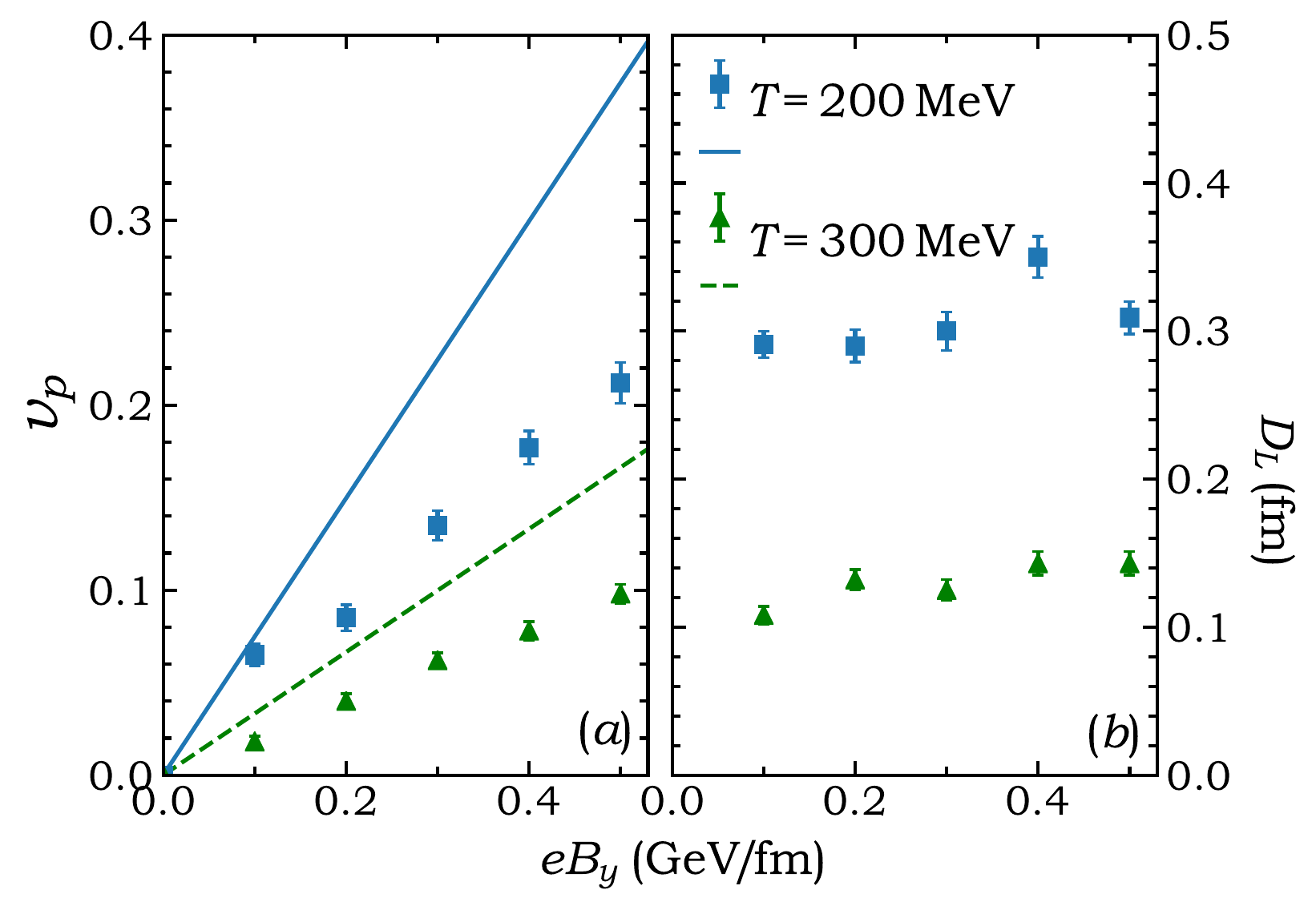}
            \caption{(Color online) Dependence of the phase velocity $v_p$ (a) and the diffusion constant $D_L$ (b) on the strength of the magnetic field at different temperatures but with the same specific shear viscosity $\eta/s=0.08\hbar$. The lines are theoretical limits from Eq. (\ref{v_p_t}) and the symbols are the simulation results.}
            \label{v_p_D_l}
    \end{figure}

    The dependence of the phase velocity and the diffusion constant of the CMW on the strength of the magnetic field at different temperatures but with the same specific shear viscosity $\eta/s=0.08\hbar$ are shown in Fig. \ref{v_p_D_l}. The phase velocity from simulations shown as symbols are smaller compared to lines from the theoretical limit of Eq. (\ref{v_p_t}), as a result of the momentum truncation. This is expected as it can be seen from Fig.~\ref{J_J5_2} that the currents are underestimated. The deviation from the theoretical limit due to the momentum truncation becomes smaller under a weaker magnetic field. It is seen in Fig.~\ref{v_p_D_l} that the phase velocity of the CMW increases with the increasing of its driving force, i.e., the strength of the magnetic field, while the diffusion constant seems to be insensitive to the strength of the magnetic field since it is dominated by the collision process. Both the phase velocity and the diffusion constant decrease with increasing temperature. Similar to that obtained from the Holographic QCD calculation~\cite{K_CMW,ss}, the increasing trend of $v_p$ with the strength of the magnetic field is reproduced, and a similar temperature dependence is observed. However, the dependence of $D_L$ on the strength of the magnetic field and the temperature seems to be different from that obtained by the Sakai-Sugimoto model \cite{K_CMW,ss}.

\begin{figure}[htb]
\includegraphics[angle=0,scale=0.5]{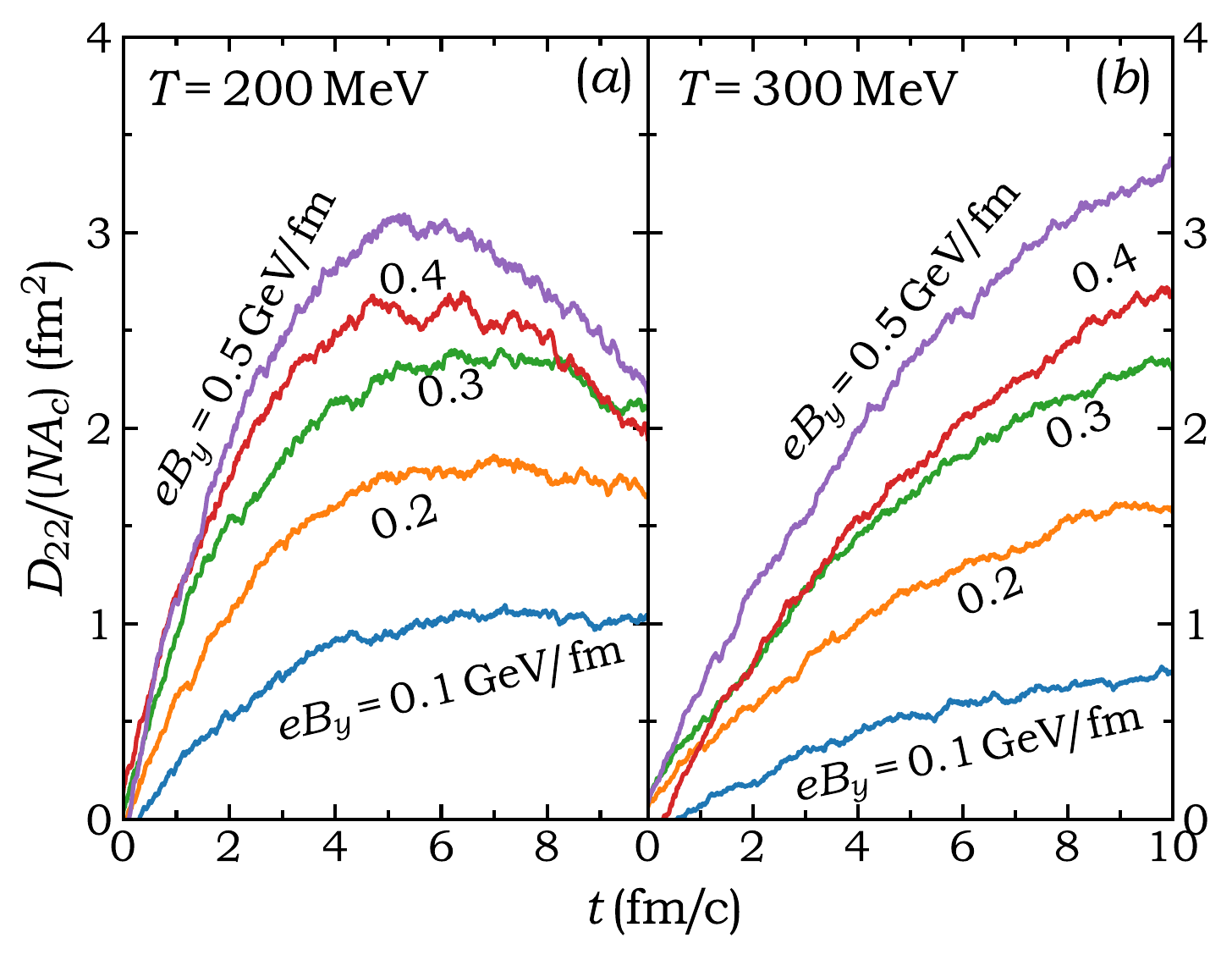}
            \caption{(Color online) Time evolution of the reduced electric quadrupole moment at different temperatures and under different strength of the magnetic field with the same specific shear viscosity $\eta/s=0.08\hbar$.}
            \label{rupole}
    \end{figure}

    The electric quadrupole moment is one of the key consequences of the CMW, and it may lead to the elliptic flow splitting between particles of oppositive charges~\cite{elec}, which can be observed experimentally~\cite{STAR15}. In relativistic heavy-ion collisions, at the initial stage the finite partonic system may have net electric charge, and the axial charge current is then generated from the CSE along the direction of the magnetic field, resulting in the similar initial density distribution as in Eq. (\ref{init}). It is thus useful to investigate how large the electric quadrupole moment can be developed and its dependence on the magnetic field and the temperature of the system. The general definition of the quadrupole moment can be expressed as
    \begin{equation}
    \mathcal{D}_{ij} = \int\rho(\vec{r})\left(3r_ir_j-r^2\delta_{ij} \right)\mathrm{d}^3 r, \label{Dint}
    \end{equation}
    where $i,j=1 \sim 3$ represents the index of the cartesian coordinate. Here we only consider the $\mathcal{D}_{22}$ component along the magnetic field, which can be expressed analytically as
    \begin{equation}
    \mathcal{D}_{22} = \frac{4A_cN}{\beta^2}e^{-D_L\beta^2t}\sin(\beta v_p t) \label{D_22}
    \end{equation}
    by substituting Eq. (\ref{r_charge}) into Eq. (\ref{Dint}), where $N$ is the total number of particles in the box. It is expected that the strength of $\mathcal{D}_{22}$ is proportional to $A_c$. It is further seen that there are competition effects on the development of $\mathcal{D}_{22}$, with the $\sin(\beta v_p t)$ term helping to develop the electric quadrupole moment while the $e^{-D_L\beta^2t}$ term decaying it. Figure~\ref{rupole} displays the time evolution of the reduced electric quadrupole moment from the box simulation by counting particles according to Eq. (\ref{Dint}) at different temperatures and under different strength of the magnetic field. Generally, the electric quadrupole moment develops faster at lower temperatures and under a stronger magnetic field. It is also seen that the electric quadrupole moment keeps on increasing at higher temperatures, while it may become saturated and decrease at lower temperatures due to the larger diffusion constant as shown in Fig.~\ref{v_p_D_l}.

    \section{SUMMARY}
    \label{sec:SUMMARY}

    In a cubic box with the periodic boundary condition under a uniform and constant external magnetic field, we have studied the chiral magnetic wave as a result of the interplay between the chiral magnetic effect and the chiral separation effect based on the chiral kinetic equations of motion. Although unavoidable truncations in the momentum space are needed in the simulation, which underestimate the strength of the chiral magnetic effect and the chiral separation effect, this doesn't hamper us from understanding properties of the chiral magnetic wave from simulations. It is found that the phase velocity of the chiral magnetic wave increases with the increasing strength of the magnetic field, while the damping of the chiral magnetic wave is insensitive to the strength of the magnetic field but more sensitive to the specific shear viscosity of the system. Both the phase velocity and the damping of the chiral magnetic wave are enhanced at lower temperatures. It is further found that the electric quadrupole moment develops faster under a stronger magnetic field or at lower temperatures. Our study serves as a baseline for further simulations of chiral anomalies in relativistic heavy-ion collisions.

\begin{acknowledgements}
We thank Chen Zhong for maintaining the high-quality performance of the computer facility, and acknowledge helpful discussions with Che Ming Ko and Xu-Guang Huang. This work was supported by the Major State Basic Research Development Program (973 Program) of China under Contract Nos. 2015CB856904 and 2014CB845401, the National Natural Science Foundation of China under Grant Nos. 11475243 and 11421505, and the Shanghai Key Laboratory of Particle Physics and Cosmology under Grant No. 15DZ2272100.
\end{acknowledgements}
~\\

\appendices
\renewcommand\thesection{APPENDIX}

\section{Determine cross section from specific shear viscosity}
\label{sec:APPENDIX}

In this appendix we briefly review the way we determine the isotropic cross section $\sigma_{22}$ from the specific shear viscosity $\eta/s$. In a thermalized medium, the shear viscosity $\eta$ can be approximately calculated from
    \begin{equation}
    \eta = \frac{4\left\langle k\right\rangle}{15\sigma_{tr}}, \label{eta}\\
    \end{equation}
where $\left\langle k\right\rangle$ is the mean momentum of particles, and $\sigma_{tr}$ is the transport cross section defined as
    \begin{equation}
    \sigma_{tr} = \int\mathrm{d}\Omega\frac{\mathrm{d}\sigma_{22}}{\mathrm{d}\Omega}\left(1-\cos^2\theta\right). \label{sigma_tr}
    \end{equation}
    $\sigma_{tr}=\frac{2}{3}\sigma_{22}$ can be obtained if $\sigma_{22}$ is an isotropic cross section. The entropy density $s$ can be written as
\begin{equation}
    s = -4N_c\int\frac{\mathrm{d}^3k}{\left(2\pi\hbar\right)^3}\left[f\ln  f-\left(1-f\right)\ln\left(1-f\right)\right],\label{sden}
\end{equation}
    where 4 represents 4 types of particles with different charge numbers and helicities, and $f\!=\!1/\left(e^{k/T}+1\right)$ is the average momentum distribution for 4 types of massless particles. Using the constraint from the elliptic flow based on the hydrodynamic calculation $\eta/s \in (0.08\hbar, 0.20\hbar)$~\cite{shear2}, and combining Eqs. (\ref{eta}-\ref{sden}), we obtain the range of the cross section depending on the temperature as
    \begin{equation}
    \sigma_{22}\in\left(1.74,\:4.36\right)\frac{\hbar^2}{T^2},
    \end{equation}
    where $\sigma_{22}$ is in fm$^2$ and $T$ is in GeV.


\begin{thebibliography}{99}
        \bibitem{nature} Q. Li {\it et al.}, Nat. Phys. {\bf 12}, 550 (2016).
        \bibitem{science} J. Xiong {\it et al.}, Science {\bf 350}, 413 (2015).
        \bibitem{PRX} X.-C. Huang {\it et al.}, Phys. Rev. X {\bf 5}, 031023 (2015).
        \bibitem{Kha16} D. E. Kharzeev, J. Liao, S. A. Voloshin, and G. Wang, Prog. Part. Nucl. Phys. \textbf{88}, 1
(2016).
        \bibitem{Hua16} X.-G. Huang, Rep. Prog. Phys. \textbf{79}, 076302 (2016).
        \bibitem{K2008} D. E. Kharzeev, L. D. McLerran, and H. J. Warringa, Nucl. Phys. A {\bf 803}, 227 (2008).
        \bibitem{Fuku1} K. Fukushima, Lect. Notes Phys. {\bf 871}, 241 (2013).
        \bibitem{K_CME} K. Fukushima, D. E. Kharzeev, and H. J. Warringa, Phys. Rev. D {\bf 78}, 074033 (2008).
        \bibitem{STAR1} B. I. Abelev {\it et al.} (STAR Collaboration), Phys. Rev. Lett. {\bf 103}, 251601 (2009).
        \bibitem{STAR2} B. I. Abelev {\it et al.} (STAR Collaboration), Phys. Rev. C {\bf 81}, 054908 (2010).
        \bibitem{ALICE1} B. I. Abelev {\it et al.} (ALICE Collaboration), Phys. Rev. Lett. {\bf 110}, 012301 (2013).
        \bibitem{CSE1} D. T. Son and A. R. Zhitnitsky, Phys. Rev. D {\bf 70}, 074018 (2004).
        \bibitem{CSE2} M. A. Metlitski and A. R. Zhitnitsky, Phys. Rev. D {\bf 72}, 045011 (2005).
        \bibitem{K_CMW} D. E. Kharzeev and H.-U. Yee, Phys. Rev. D {\bf 83}, 085007 (2011).
        \bibitem{elec} Y. Burnier, D. E. Kharzeev, J.-F. Liao, and H.-U. Yee, Phys. Rev. Lett. {\bf 107}, 052303 (2011).
        \bibitem{STAR15} L. Adamczyk {\it et al.} (STAR Collaboration), Phys. Rev. Lett. \textbf{114}, 252302 (2015).

        \bibitem{CMS1} V. Khachatryan {\it et al.} (CMS Collaboration), Phys. Rev. Lett. {\bf 118}, 122301 (2017).
        \bibitem{CMS2} A. M. Sirunyan {\it et al.} (CMS Collaboration), Phys. Rev. C {\bf 97}, 044912 (2018).
        \bibitem{back} F.-Q. Wang and J. Zhao, Phys. Rev. C {\bf 95}, 051901(R) (2017).
        \bibitem{Xu12} J. Xu, L. W. Chen, C. M. Ko, and Z. W. Lin, Phys. Rev. C \textbf{85}, 041901(R) (2012).
        \bibitem{Xu14} J. Xu, T. Song, C. M. Ko, and F. Li, Phys. Rev. Lett. \textbf{112}, 012301 (2014).
        \bibitem{Liu16} H. Liu, J. Xu, L. W. Chen, and K. J. Sun, Phys. Rev. D \textbf{94}, 065032 (2016).
        \bibitem{Zha17} J. Zhao, H. L. Li, and F. Q. Wang, arXiv: 1705.05410 [nucl-ex].
        \bibitem{Mag18} N. Magdy {\it et al.}, Phys. Rev. C \textbf{97}, 061901 (2018).
        \bibitem{Xu18} H. J. Xu {\it et al.}, Chin. Phys. C {\bf 42}, 084103 (2018).
        \bibitem{rhic_iso} \url{https://drupal.star.bnl.gov/STAR/files/STAR_BUR_Run1617_v18.pdf}.
        \bibitem{Hydro1} M. V. Isachenkov and A. V. Sadofyev, Phys. Lett. B {\bf 697}, 404 (2011).
        \bibitem{Hydro2} D. T. Son and P. Sur\'owka, Phys. Rev. Lett. {\bf 103}, 191601 (2009).
        \bibitem{Liao} S.-Z. Shi, Y. Jiang, E. Lilleskov, and J.-F. Liao, Ann. of Phys. {\bf 394}, 50 (2018).
        \bibitem{CKM1} M. A. Stephanov and Y. Yin, Phys. Rev. Lett. {\bf 109}, 162001 (2012).
        \bibitem{CKM3} J.-W. Chen, S. Pu, Q. Wang, and X.-N. Wang, Phys. Rev. Lett. {\bf 110}, 262301 (2013).
        \bibitem{CKM4} D. T. Son and N. Yamamoto, Phys. Rev. D {\bf 87}, 085016 (2013).
        \bibitem{sun1} Y.-F. Sun, C. M. Ko, and F. Li, Phys. Rev. C {\bf 94}, 045204 (2016).
        \bibitem{sun2} Y.-F. Sun and C. M. Ko, Phys. Rev. C {\bf 95}, 034909 (2017).
        \bibitem{stochastic} Z. Xu and C. Greiner, Phys. Rev. C {\bf 71}, 064901 (2005).
        \bibitem{Dem09} N. Demir and S. A. Bass, Phys. Rev. Lett. \textbf{102}, 172302 (2009).
        \bibitem{Zha18} Z. Zhang and C. M. Ko, Phys. Rev. C \textbf{97}, 014610 (2018).
        \bibitem{box2} Y.-X. Zhang {\it et al.}, Phys. Rev. C {\bf 97}, 034625 (2018).
        \bibitem{huangy} E. van der Bijl and R. A. Duine, Phys. Rev. Lett. {\bf 107}, 195302 (2011).
        \bibitem{Huang} X.-G. Huang, Sci. Rep. {\bf 6}, 20601 (2016).
        \bibitem{Berry1984} M. V. Berry, Proc. R. Soc. Lond. A {\bf 392}, 45 (1984).
        \bibitem{gc} D. Xiao, J. R. Shi, and Q. Niu, Phys. Rev. Lett. \textbf{95}, 134204 (2005).
        \bibitem{JHEP} G. M. Newman, J. High Energy Phys. {\bf 01}, 158 (2006).
        \bibitem{ss} T. Sakai and S. Sugimoto, Prog. Theor. Phys. {\bf 113}, 843 (2005); {\bf 114}, 1083 (2005).
        \bibitem{shear2} H.-C. Song, S. A. Bass, U. Heinz, T. Hirano, and C. Shen, Phys. Rev. Lett. {\bf 106}, 192301 (2011).
    \end{thebibliography}
\end{document}